\documentclass{ptephy_v1}

\usepackage{amsmath,amssymb,amsthm,graphicx,physics}
\def\e{\mathrm{e}}
\def\d{\partial}

\newtheorem*{thm*}{Theorem}

\begin{document}

\title{Distributions of consecutive level spacings of
circular unitary ensemble and their ratio:\\
finite-size corrections and Riemann $\zeta$ zeros}

\author{\name{\fname{Shinsuke} \surname{M. Nishigaki}}{\ast}} 

\address{\affil{~}{
Graduate School of Natural Science and Technology,
Shimane University, Matsue 690-8504, Japan}
\email{nishigaki@riko.shimane-u.ac.jp}}

\begin{abstract}
We compute the joint distribution of two consecutive eigenphase spacings and their ratio for
Haar-distributed $\mathrm{U}(N)$ matrices (the circular unitary ensemble)
using our framework for J\'{a}nossy densities in random matrix theory,
formulated via the Tracy-Widom system of nonlinear PDEs.
Our result shows that the leading finite-$N$
correction in the gap-ratio distribution relative to
the universal sine-kernel limit is of $\mathcal{O}(N^{-4})$,
reflecting a nontrivial cancellation of the $\mathcal{O}(N^{-2})$ part
present in the joint distributions of consecutive spacings.
This finding suggests the potential to extract subtle finite-size corrections
from the energy spectra of quantum-chaotic systems and explains why 
the deviation of the gap-ratio distribution of the Riemann zeta zeros
$\{1/2+i\gamma_n\}, \gamma_n\approx T\gg1$ from the sine-kernel prediction
scales as $\left(\log(T/2\pi)\right)^{-3}$.
\end{abstract}

\subjectindex{A10,A13,A32,B83,B86}

\maketitle

\section{Introduction}
Circular ensembles of random matrices, i.e.~ensembles of Haar-distributed 
unitary matrices of large rank that parametrize compact Riemannian symmetric spaces,
were first introduced by Dyson \cite{Dyson:1962} as idealized models
of the time-evolution operators $\e^{-itH}$ of complex quantum systems.
Decades later, it was established that the circular unitary $\mathrm{U}(N)$ ensemble
(hereafter denoted as CUE$_N$)
exhibits finite-$N$ properties common to those of the Riemann zeta function
$\zeta(1/2+iT)$ under identification 
$N\propto\log ({T}/{2\pi})$ 
\cite{Keating:2000,Bogomolny:2006,Forrester:2015,Bornemann:2017}.
As the zeta function is arguably regarded as an ``ideal'' quantum chaotic system,
circular ensembles have come to be recognized as sources of ``canonical" finite-size
corrections to the universality of random matrix theory in the large-$N$ limit,
prompting a wave of researches into the deviations of $\order{N^{-2}}$ 
(see \cite{Forrester:2025} and references therein).

On the other hand, recent studies of quantum many-body and chaotic systems
have popularized the gap-ratio distribution
$P_{\mathrm{r}}(r)$, i.e. the distribution of the ratios of
two consecutive energy level spacings $r=(E_{n+1}-E_n)/(E_n-E_{n-1})$,
as a leading measure of quantum chaoticity
\cite{Oganesyan:2007,Atas:2013,Atas:2013b}.
The primary reason for this popularity is
that the gap ratio $r$ is insensitive to the global density of states (DOS) $\rho(E)$,
which is system-specific and can grow rapidly with energy
even within a narrow spectral window\footnote{
Such DOS's of physical relevance include
the mobility edge of QCD Dirac spectra at $T>T_{\mathrm{cr}}$ \cite{Nishigaki:2013}.},
unlike the conventional level spacing $s=\rho(E_n)(E_{n+1}-E_n)$.
In light of these two trends this article aims to present analytical expressions for
two spectral-statistical distributions of CUE$_N$,
the gap-ratio distribution and a related quantity:
the distribution $P_{\mathrm{nn}}(t)$ of the nearest-neighbor level spacing
$t=\rho(E_n)\min(E_{n+1}-E_n, E_n-E_{n-1})$ \cite{Forrester:1996}.
These results are derived using a framework developed by the author
in earlier works \cite{Nishigaki:2021,Nishigaki:2024},
which treated the Airy, Bessel, and sine kernels as illustrative examples.

This article is structured as follows.
In Section 2, we list key properties of the determinantal point process (DPP)
governing the eigenphases of CUE$_N$, with a focus on its J\'{a}nossy density
defined as the probability that a given spectral interval contains no eigenphase
except one at a specified location.
Section 3 revisits the Tracy-Widom (TW) formalism \cite{Tracy:1994c}
for {\em exponential variants} of 
integrable kernels and presents a system of partial differential equations (PDEs)
that governs the Fredholm determinant representing the J\'{a}nossy density of CUE$_N$.
In Section 4, we derive  the nearest-neighbor spacing distribution $P_{\mathrm{nn}}(t)$,
the joint distribution of two consecutive spacings  $P_{\mathrm{c}}(a,b)$, 
and the gap-ratio distribution $P_{\mathrm{r}}(r)$ from the J\'{a}nossy density,
and examine their deviations from the sine-kernel limit $N\to\infty$.
Finally, in Section 5, we present a numerical study of how the joint distribution of
consecutive spacings and the gap-ratio distribution of the Riemann zeta zeros
$\{1/2+ i \gamma_n\}$ at increasing heights $\gamma_n\approx T\gg 1$ deviate from
sine-kernel predictions, 
and explain the $\left(\log(T/2\pi)\right)^{-3}$-scaling observed in the latter.

\section{J\'{a}nossy density of CUE}
Let $\{\e^{ix_1},\ldots, \e^{ix_N}\}$
be the set of eigenvalues of a $\mathrm{U}(N)$ matrix $U$.
If $U$ is Haar-distributed, the probability distribution of $\{x_i\}$
is a DPP governed by a kernel of the form \cite{Dyson:1962}
\begin{align}
K(x,y)=\frac{1}{2\pi}\frac{\sin (N(x-y)/2)}{\sin ((x-y)/2)}.
\nonumber
\end{align}
It enjoys the projective and normalization properties
$\int_{-\pi}^\pi dy\,K(x,y)K(y,z)=K(x,z)$,
$\int_{-\pi}^\pi dx\,K(x,x)=N$,
requisite for a DPP. 
In this section we employ two-component notations:
\begin{align}
K(x,y)=\frac{{}^t\Phi(x) J \Phi(y)}{\e^{ix}-\e^{iy}},~~
\Phi(x)=
\left[
\begin{array}{c}
\varphi(x)\\
\psi(x)
\end{array}
\right]
=\frac{\e^{ix/2}}{\sqrt{2\pi}}
\left[
\begin{array}{c}
\e^{iNx/2}\\
\e^{-iNx/2}
\end{array}
\right],~~
J=
\left[
\begin{array}{cc}
0 & 1\\
-1 & 0
\end{array}
\right].
\label{K_CUE}
\end{align}
After unfolding $x\mapsto (2/N)x$ to normalize the mean level spacing to $\pi$,
this CUE$_N$ kernel converges to the universal sine kernel in the large-$N$ limit
with $x-y$ held fixed,
\begin{align}
K(x,y)|_{\mathrm{unfold.}}
&=\frac{\sin (x-y)}{\pi N \sin ((x-y)/N)}
\nonumber\\
&=\frac{\sin(x-y)}{\pi(x-y)}+
\frac{(x-y) \sin (x-y)}{6 \pi \,N^2}+
\frac{7(x-y)^3 \sin (x-y)}{360 \pi \,N^4}+
\mathcal{O}\Bigl(\frac{1}{N^6}\Bigr).
\label{Ksin}
\end{align}

Given that the $p$-point correlation function ${R}_p(x_1,\ldots,x_p)$ of a DPP
is expressed as a $p\times p$ determinant $\det[K(x_i, x_j)]_{i,j=1}^p$ of the kernel, 
the conditional $p$-point correlation function
$\tilde{R}_{p|1}(x_1,\ldots,x_p|s)={R}_{p+1}(x_1,\ldots,x_p,s)/{R}_1(s)$
where one of the $x$'s is fixed at a designated value $s$
also retains a determinantal form $\det\bigl[\tilde{K}(x_i, x_j)\bigr]_{i,j=1}^p$
governed by a modified kernel
\begin{align}
\tilde{K}(x,y)=K(x,y)-K(x,s)K(s,s)^{-1}K(s,y).
\label{Ktransf}
\end{align}
In the case of CUE$_N$, this modified kernel was originally presented in [5, Eq.\,(44)].
It again enjoys the properties
$\int_{-\pi}^\pi dy\,\tilde{K}(x,y)\tilde{K}(y,z)=\tilde{K}(x,z)$,
$\int_{-\pi}^\pi dx\,\tilde{K}(x,x)=N-1$.

By translational invariance under the shift $x_j\mapsto x_j-s$ for all $j$,
the parameter $s$ can be set to zero without loss of generality.
The transformation between the kernels (\ref{Ktransf}) is associated with an 
$\mathrm{SL}(2,\mathbb{C})$ gauge transformation 
$\tilde{\Phi}(x)=U(x)\Phi(x)$ between the two-component functions \cite{Nishigaki:2021} with
(in what follows $\mathbb{I}$ stands for the identity matrix or operator)
\[
U(x)^{\pm1}=
\mathbb{I}\pm\frac{\Phi(0)\,{}^t\Phi(0)J}{K(0,0)(\e^{ix}-1)},~~
\det U(x)=1,
\]
and the modified kernel (\ref{Ktransf}) assumes the same ``integrable" form as (\ref{K_CUE}):
\begin{align}
&\tilde{K}(x,y)=\frac{{}^t \tilde{\Phi}(x) J \tilde{\Phi}(y)}{\e^{ix}-\e^{iy}},
\nonumber\\
&\tilde{\Phi}(x)=
\left[
\begin{array}{c}
\tilde{\varphi}(x)\\
\tilde{\psi}(x)
\end{array}
\right]
=\frac{\e^{ix/2}}{\sqrt{2\pi}}
\left[
\begin{array}{c}
{\displaystyle \e^{iNx/2}-\frac{\e^{iNx/2}-\e^{-iNx/2}}{N(\e^{ix}-1)}}\\
{\displaystyle \e^{-iNx/2}-\frac{\e^{iNx/2}-\e^{-iNx/2}}{N(\e^{ix}-1)}}
\end{array}
\right].
\label{Ktil}
\end{align}

In their seminal work \cite{Tracy:1994c}, Tracy and Widom established the following theorem:
\begin{thm*}[Tracy-Widom, Sect.\,II.E]
Consider an involution kernel of the form $(b\in \mathbb{C})$
\begin{align}
K(x,y)=
\frac{\varphi(x)\psi(y)-\psi(x)\varphi(y)}{\e^{bx}-\e^{by}}=
\frac{{}^t\Phi(x) J \Phi(y)}{\e^{bx}-\e^{by}}.
\label{CD_exp}
\end{align}
If the two-component function $\Phi(x)={}^t (\varphi(x), \psi(x))$
satisfies a first-order linear DE
\begin{align}
&\left(\frac{d}{dx} -\frac{b}{2}\right)\Phi(x) = \Omega(x) \Phi(x),~
\Omega(x)=
\frac{1}{m(x)} 
\left[
\begin{array}{rr}
A(x)& B(x)\\
-C(x) & -A(x)
\end{array}
\right]
\nonumber\\
&~~~~~~~~~~~~~
\mbox{where}~m, A, B, C~\mbox{are polynomials in}~ \e^{b x},
\label{integrable}
\end{align}
then the Fredholm determinant $\det(\mathbb{I}-\mathbb{K}_I)$ of the integration operator
$\mathbb{K}_I$, with kernel ${K}(x,y)$ acting on the space of $L^2(I)$ functions over
a union of intervals $I\subset\mathbb{R}$,
is determined by a system of PDEs with respect to the end-points of $I$.
\end{thm*}
In our case at hand, the original $\Phi(x)$ satisfies (\ref{integrable}) with
$b=i$, $m=i$, $A=-N/2$, $B=C=0$.
Accordingly, the gauge-transformed function $\tilde{\Phi}(x)$
necessarily satisfies the TW criteria (\ref{CD_exp}), (\ref{integrable})
associated with a gauge-transformed $\mathfrak{sl}(2,\mathbb{C})$ connection
$\tilde{\Omega}(x)$
\cite{Nishigaki:2021}:
\begin{align}
&\Omega(x)\mapsto \tilde{\Omega}(x)=
U(x) \Omega(x) U(x)^{-1}-U(x)\frac{d}{dx}U(x)^{-1}
=\frac{1}{\tilde{m}(x)}
\left[
\begin{array}{rr}
\tilde{A}(x)& \tilde{B}(x)\\
-\tilde{C}(x) & -\tilde{A}(x)
\end{array}
\right],
\nonumber\\
&\tilde{m}(x)=i(1-\e^{i x}),~~
\tilde{A}(x)= -\frac{N}{2}(1-\e^{ix})+\frac{1}{N},~~
\tilde{B}= -1-\frac{1}{N},~~
\tilde{C}= 1-\frac{1}{N}.
\label{coeffs}
\end{align}
Subsequently we shall need the coefficients of these polynomials in $\e^{ix}$, denoted as
\begin{align}
\mu_0=i,~
\mu_1=-i,~
\alpha_0=-\frac{N}{2}+\frac{1}{N},~
\alpha_1=\frac{N}{2},~
\beta_0=-1-\frac{1}{N},~
\gamma_0=1-\frac{1}{N}.
\label{coeffs}
\end{align}
We now focus on the single-interval case $I=[a_1, a_2]$ with
$a_1< 0< a_2$ and $a_2-a_1<2\pi$ so that the ordered triple 
$(a_1, 0, a_2)$ represents three consecutive eigenphases,
and denote 
$\tilde{\mathbb{K}}_I \doteq \tilde{K}(x,y)\chi_I(y)$.\footnote{
Tracy-Widom's notation $\mathbb{L} \doteq L(x,y)$ means 
that the integration operator $\mathbb{L}$ has kernel $L(x,y)$.}
Then, by the Gaudin-Mehta theorem \cite{Forrester:2010},
the conditional probability $\tilde{J}_1(0;[a_1,a_2])$ that the interval $[a_1, a_2]$
contains no eigenphase other than the one preconditioned at $x=0$
is given by the Fredholm deterninant of $\tilde{\mathbb{K}}_I$:
\begin{align}
\tilde{J}_1(0;[a_1,a_2])&=
\det(\mathbb{I}-\tilde{\mathbb{K}}_I)=
\exp\Bigl(-\sum_{n\geq 1}\frac1n\mathrm{tr}\,\tilde{\mathbb{K}}_I^n\Bigr)
\nonumber\\
&=\exp\left(
-\int_I dx\,\tilde{K}(x,x)
-\frac12 \int\!\!\!\!\int_I dx dy\,\tilde{K}(x,y)\tilde{K}(y,x)
-\cdots \right),
\label{FredholmDet}
\end{align}
and can be computed using the TW framework, as detailed in the following section.
The J\'{a}nossy density (measure) ${J}_1(0;[a_1,a_2])dx$, defined as
the unconditional probability that the interval $[a_1, a_2]$
contains no eigenphase except for one located in $[0, dx]$, is given by
$\tilde{J}_1(0;[a_1,a_2])dx$ multiplied by the DOS,
which is $K(0,0)={N}/{2\pi}$ in our case.

\section{Tracy-Widom theory revisited}
\subsection{Preliminary}
Having quoted the core content of {\bf Theorem} from \cite{Tracy:1994c},
we find it worthwhile to revisit the case where {\em $m(x)$ is linear in $\e^{bx}$}
for the sake of this article's self-consistency. 
In \cite{Tracy:1994c},
not all terms in the corresponding system of PDEs
for this ``exponential variant'' case \cite{Nishigaki:1998,Nishigaki:1999}
with $m(x)\neq$ const. were explicitly spelt out,
and an incorrect inclusion of terms was presented.

We employ notation $\varphi_\ell(x)=\e^{\ell bx} \varphi(x)$,
$\psi_\ell(x)=\e^{\ell bx} \psi(x)$ with non-negative integer indices $\ell$.
Quantities involved in the TW system for $I=[a_1,a_2]$ are ($j,k=1$ or $2$):
\begin{align}
R_{jk}&=(\mathbb{I}-\mathbb{K}_I)^{-1}{{K}}(a_j, a_k)=R_{kj}
\nonumber\\
&={K}(a_j, a_k)+
\int_I dx \, {K}(a_j, x){K}(x, a_k)+
\int\!\!\!\!\int_I dx dy\, {K}(a_j, x){K}(x, y){K}(y, a_k)+\cdots,
\nonumber\\
q_{\ell j}
&=(\mathbb{I}-\mathbb{K}_I)^{-1} {\varphi}_\ell(a_j)
\nonumber\\
&={\varphi}_\ell(a_j)+
\int_I dx \, {K}(a_j, x)\varphi_\ell(x)+
\int\!\!\!\!\int_I dx dy\, {K}(a_j, x){K}(x, y)\varphi_\ell(y)+\cdots,
\nonumber\\
p_{\ell j}&=(\mathbb{I}-\mathbb{K}_I)^{-1}  {\psi}_\ell(a_j),
\nonumber\\  
u_\ell&=
\int_I dx\,  {\varphi}(x) \, (\mathbb{I}-\mathbb{K}_I)^{-1} {\varphi}_\ell(x),~~
w_\ell=\int_I dx\,{\psi}(x) \,  (\mathbb{I}-\mathbb{K}_I)^{-1} {\psi}_\ell(x),
\nonumber\\
v_\ell&=\int_I dx\,{\psi}(x)  \, (\mathbb{I}-\mathbb{K}_I)^{-1} {\varphi}_\ell(x),~~
\tilde{v}_\ell=\int_I dx\, {\varphi}(x)  \, (\mathbb{I}-\mathbb{K}_I)^{-1} {\psi}_\ell(x).
\label{vars}
\end{align}
All these quantities are regarded as functions of the two endpoints $a_1$ and $a_2$ of $I$.
For brevity, we denote $q_j=q_{0j}$, $u=u_0$, etc. for the quantities
with index $\ell=0$.

\subsection{Universal equations}
Due to the property of the indicator function $\chi_{I}(y)=\Theta(y-a_1)\Theta(a_2-y)$, 
\begin{align}
\frac{\partial }{\partial a_k}\chi_{I}(y) =(-1)^k \delta(y-a_k),
\nonumber
\end{align}
the Fredholm determinant $\det(\mathbb{I}-\mathbb{K}_I)$
is determined by the diagonal resolvent $R_{kk}$ via
\begin{align}
\frac{\d}{\d a_k}\log\det(\mathbb{I}-\mathbb{K}_I)=-
\mathrm{tr}\left((\mathbb{I}-\mathbb{K}_I)^{-1}\frac{\partial\mathbb{K}_I}{\partial a_k}\right)
=-(-1)^{k} R_{kk}.
\label{logdet}
\end{align}
By the same token, differentiation of 
$q_j, p_j, u_\ell, v_\ell, \tilde{v}_\ell, w_\ell$ in (\ref{vars})
with respect to $a_{k (\neq j)}$ leads to
[in what follows, the pair of indices $(j,k)$ takes the values either (1,2) or (2,1)]:
\begin{align}
\frac{\d q_j}{\d a_k}&=(-1)^k R_{jk} q_k,~
\frac{\d p_j}{\d a_k}=(-1)^k R_{jk} p_k~,
\label{vars_der_k1}\\
\frac{\d u_\ell}{\d a_k}&=(-1)^k q_{\ell k} q_k,~
\frac{\d w_\ell}{\d a_k} =(-1)^k p_{\ell k} p_k,~
\frac{\d v_\ell}{\d a_k} =(-1)^k p_{\ell k} q_k,~
\frac{\d \tilde{v}_\ell}{\d a_k} =(-1)^k q_{\ell k} p_k.
\label{vars_der_k2}
\end{align}
On the other hand, differentiation of $q_j$ or $p_j$ with respect to $a_{j}$ 
involves additional terms
$\bigl(D(\mathbb{I}-\mathbb{K}_I)^{-1} {\varphi}(x)\bigr)|_{x=a_j}$, etc.:
\begin{align}
\frac{\d q_j}{\d a_j}&=(-1)^j R_{jj} q_j
+D(\mathbb{I}-\mathbb{K}_I)^{-1} \varphi(a_j),~
\label{vars_del_j1}\\
\frac{\d p_j}{\d a_j}&=(-1)^j R_{jj} p_j
+D(\mathbb{I}-\mathbb{K}_I)^{-1} \psi(a_j),
\label{vars_del_j2}
\end{align}
where $D$ denotes the differentiation operator with respect to the independent variable $x$.

Let $X$ denote the multiplication operator by $\e^{bx}$.
Eq.\,(\ref{CD_exp}) means that the commutator of $X$ and $\mathbb{K}_I$ is given by
\[
[X, \mathbb{K}_I]\doteq
\e^{bx} K(x,y)\chi_I(y) - K(x,y)\chi_I(y) \e^{by}=
(\varphi(x)\psi(y)-\psi(x)\varphi(y))\chi_I(y).
\]
It immediately leads to:
\begin{align}
&[X, (\mathbb{I}-\mathbb{K}_I)^{-1}]=
(\mathbb{I}-\mathbb{K}_I)^{-1}[X,\mathbb{K}_I](\mathbb{I}-\mathbb{K}_I)^{-1} 
\nonumber\\
&\doteq \left(
(\mathbb{I}-\mathbb{K}_I)^{-1} \varphi(x) \, 
(\mathbb{I}-{}^t \mathbb{K}_I)^{-1} \psi(y) -
(\mathbb{I}-\mathbb{K}_I)^{-1} \psi(x) \, 
(\mathbb{I}-{}^t \mathbb{K}_I)^{-1} \varphi(y)
\right)\chi_I(y).
\label{MK1}
\end{align}
However, it also holds from an identity
$(\mathbb{I}-\mathbb{K}_I)^{-1}=\mathbb{I}+(\mathbb{I}-\mathbb{K}_I)^{-1}\mathbb{K}_I$ that
\begin{align}
[X, (\mathbb{I}-\mathbb{K}_I)^{-1}]\doteq
(\e^{bx}-\e^{by})(\mathbb{I}-\mathbb{K}_I)^{-1}K(x,y)\chi_I(y).
\label{MK2}
\end{align}
By equating (\ref{MK1}) and (\ref{MK2}) and setting $x\to a_j$ and $y\to a_{k(\neq j)}$
while $x,y\in I$,
one obtains
\begin{align}
R_{jk}=\frac{q_j p_k-p_j q_k}{\e^{b a_j}-\e^{b a_k}},
\label{Rjk}
\end{align}
whereas setting $x\to a_j$ and then $y\to a_j$ leads to, by l'H\^{o}pital's rule,
\begin{align}
R_{jj}=\frac{1}{b\,\e^{b a_j}}\left(p_j\frac{\d q_j}{\d a_j}-q_j\frac{\d p_j}{\d a_j}\right).
\label{Rkk}
\end{align}
Similarly, the commutator of $X^\ell$ and $\mathbb{K}_I$
\begin{align}
[X^\ell, \mathbb{K}_I]&\doteq 
\frac{\e^{\ell b x}-\e^{\ell b y}}{\e^{b x}-\e^{b y}}
(\varphi(x)\psi(y)-\psi(x)\varphi(y))\chi_I(y)
\nonumber\\
&=\sum_{{n+n'=\ell-1 \atop n, n'\geq 0}}
(\varphi_n(x)\psi_{n'}(y)-\psi_n(x)\varphi_{n'}(y))\chi_I(y)
\nonumber
\end{align}
leads to:
\begin{align}
&[X^\ell , (\mathbb{I}-\mathbb{K}_I)^{-1}]
\doteq 
\label{Xl}\\
&\sum_{{n+n'=\ell-1 \atop n, n'\geq 0}}
\left(
(\mathbb{I}-\mathbb{K}_I)^{-1} \varphi_n(x) \, (\mathbb{I}-{}^t \mathbb{K}_I)^{-1} \psi_{n'}(y)-
(\mathbb{I}-\mathbb{K}_I)^{-1} \psi_n(x) \, (\mathbb{I}-{}^t \mathbb{K}_I)^{-1} \varphi_{n'}(y)
\right)\chi_I(y).
\nonumber
\end{align}
Applying (\ref{Xl}) to $\varphi(x)$ or $\psi(x)$ and then setting $x\to a_j, x\in I$ gives
\begin{align}
\e^{\ell b a_j} q_j-q_{\ell j}
&=\sum_{{n+n'=\ell-1 \atop n, n'\geq 0}}(v_n q_{n' j} - u_n p_{n'j}),
\label{qlj}\\
\e^{\ell b a_j} p_j-p_{\ell j}
&=\sum_{{n+n'=\ell-1 \atop n, n'\geq 0}}(w_n q_{n' j} - \tilde{v}_n p_{n'j}).
\label{plj}
\end{align}
By recursively applying (\ref{qlj}) and (\ref{plj}),
$q_{\ell j}$ and $p_{\ell j}$ for any $\ell\geq 1$ can be expressed
in terms of $q_{j}$ and $p_{j}$ at the cost of introducing
$u_n$, $w_n$, $v_n$, $\tilde{v}_n$ for $n\leq \ell-1$.
In turn, the $a_j$-dependence of the latter quantities, Eq.\,(\ref{vars_der_k2}),
is controlled by $q_{\ell' j}$ and $p_{\ell' j}$ for $\ell'\leq n$.
Thus, the closure of the PDE system depends on the $a_j$-dependence of $q_j$ and $p_j$,
(\ref{vars_del_j1}) and (\ref{vars_del_j2}).

\subsection{Nonuniversal equations}
All operations up to this point are universal.
We now incorporate nonuniversal (kernel-specific) information
into the second terms on the right-hand sides of Eqs.\,(\ref{vars_del_j1}) and (\ref{vars_del_j2}).
To this end, let $M$ denote the multiplication operator by $m(x)=\mu_0+\mu_1\e^{bx}$
and consider the commutator:
\begin{align}
&[MD, \mathbb{K}_I]\doteq 
\left( m(x)\frac{\d}{\d x}+m(y)\frac{\d}{\d y}+m'(y)\right)K(x,y)\chi_I(y)
\label{MDKI1}\\
&=\left\{\left( m(x)\frac{\d}{\d x}+m(y)\frac{\d}{\d y}+\mu_1 b\,\e^{by}\right)K(x,y)\right\}
\chi_I(y) -\sum_{k=1,2} (-1)^k m(a_k)K(x,a_k)\delta(y-a_k).
\nonumber
\end{align}
On the other hand, Eq.\,(\ref{integrable}) leads to 
\begin{align}
m(x)\frac{\d K(x,y)}{\d x} =&
\frac{
A(x)(\varphi(x)\psi(y)+\psi(x)\varphi(y))+
B(x)\psi(x)\psi(y)+
C(x)\varphi(x)\varphi(y)}{\e^{bx}-\e^{by}}
\nonumber\\
&-\frac{b}{2}m(x)\frac{\e^{bx}+\e^{by}}{\e^{bx}-\e^{by}} K(x,y)
\nonumber
\end{align}
and the above equation with $x\leftrightarrow y$ exchanged.
Accordingly it follows that
\begin{align}
&\left(
m(x)\frac{\d}{\d x} +
m(y)\frac{\d}{\d y} 
+\frac{b}{2}\mu_1
(\e^{bx}+\e^{by})
\right) K(x,y)
\label{MDKI2}\\
=&\frac{A(x)-A(y)}{\e^{bx}-\e^{by}}(\varphi(x)\psi(y)+\psi(x)\varphi(y))+
\frac{B(x)-B(y)}{\e^{bx}-\e^{by}}\psi(x)\psi(y)+
\frac{C(x)-C(y)}{\e^{bx}-\e^{by}}\varphi(x)\varphi(y).
\nonumber
\end{align}
Substitution of (\ref{MDKI2}) into (\ref{MDKI1}) gives
\begin{align}
&[MD, \mathbb{K}_I]
\nonumber\\
&\doteq \Bigl\{
\frac{A(x)-A(y)}{\e^{bx}-\e^{by}}(\varphi(x)\psi(y)+\psi(x)\varphi(y))+
\frac{B(x)-B(y)}{\e^{bx}-\e^{by}}\psi(x)\psi(y)+
\frac{C(x)-C(y)}{\e^{bx}-\e^{by}}\varphi(x)\varphi(y)
\nonumber\\
&-\frac{b}{2}\mu_1(\varphi(x)\psi(y)-\psi(x)\varphi(y))
\Bigr\}\chi_I(y)
-\sum_{k} (-1)^k m(a_k)K(x,a_k)\delta(y-a_k)
\nonumber\\
&=\Bigl\{
\sum_{\ell, \ell'\geq 0}\alpha_{\ell+\ell'+1}
\bigl(
(\varphi_\ell (x)\psi_{\ell'}(y)+\psi_\ell(x)\varphi_{\ell'}(y))+
\beta_{\ell+\ell'+1}\psi_\ell(x) \psi_{\ell'}(y)+
\gamma_{\ell+\ell'+1}\varphi_\ell(x) \varphi_{\ell'}(y)
\bigr)
\nonumber\\
&-\frac{b}{2}\mu_1(\varphi(x)\psi(y)-\psi(x)\varphi(y))
\Bigr\}\chi_I(y)
-\sum_{k} (-1)^k m(a_k)K(x,a_k)\delta(y-a_k).
\label{MDKI4}
\end{align}
In the last equality, the factors $\frac{A(x)-A(y)}{\e^{bx}-\e^{by}}$, etc.{}
are expanded in polynomials in $\e^{bx}$ and $\e^{by}$:
\begin{align}
A(x)=\sum_{\ell\geq 0}\alpha_\ell \e^{\ell bx}
~\Rightarrow~
\frac{A(x)-A(y)}{\e^{bx}-\e^{by}}= 
\sum_{\ell, \ell'\geq 0}\alpha_{\ell+\ell'+1}\e^{\ell bx} \e^{\ell' by},\ \mbox{etc.}
\nonumber
\end{align}
We now apply the commutator
\[
[MD, (\mathbb{I}-\mathbb{K}_I)^{-1}]=(\mathbb{I}-\mathbb{K}_I)^{-1}[MD,
 \mathbb{K}_I](\mathbb{I}-\mathbb{K}_I)^{-1}
\]
to $\varphi(x)$ and  $\psi(x)$, and then set $x\to a_j, x\in I$.
The left-hand sides are equal to, due to (\ref{integrable}),
\begin{align}
&MD(\mathbb{I}-\mathbb{K}_I)^{-1}\varphi(a_j)-(\mathbb{I}-\mathbb{K}_I)^{-1}MD\varphi(a_j)
\nonumber\\
&=m(a_j) D(\mathbb{I}-\mathbb{K}_I)^{-1}\varphi(a_j)
-\sum_{\ell\geq 0}\Bigl(
\alpha_\ell q_{\ell j}+\beta_\ell p_{\ell j}+\frac{b}{2}\mu_\ell q_{\ell j}
\Bigr),
\label{ls1}\\
&MD(\mathbb{I}-\mathbb{K}_I)^{-1}\psi(a_j)-(\mathbb{I}-\mathbb{K}_I)^{-1}MD\psi(a_j)
\nonumber\\
&=m(a_j) D(\mathbb{I}-\mathbb{K}_I)^{-1}\psi(a_j)
+\sum_{\ell\geq 0}\Bigl(
\gamma_\ell q_{\ell j}+\alpha_\ell p_{\ell j}-\frac{b}{2}\mu_\ell p_{\ell j}
\Bigr),
\label{ls2}
\end{align}
while the right-hand sides are, by virtue of (\ref{MDKI4}),
\begin{align}
&\sum_{\ell, \ell'\geq 0}
\bigl(
\alpha_{\ell+\ell'+1} (q_{\ell j} v_{\ell'}+p_{\ell j} u_{\ell'})+
\beta_{\ell+\ell'+1}p_{\ell j}v_{\ell'}+
\gamma_{\ell+\ell'+1}q_{\ell j} u_{\ell'}
\bigr)
-\frac{b}{2}\mu_1(q_j v-p_j u)
\nonumber\\
&-\sum_{k} (-1)^k m(a_k)R_{jk}q_k,
\label{rs1}\\
&\sum_{\ell, \ell'\geq 0}
\bigl(
\alpha_{\ell+\ell'+1} (q_{\ell j} w_{\ell'}+p_{\ell j} \tilde{v}_{\ell'})+
\beta_{\ell+\ell'+1}p_{\ell j}w_{\ell'}+
\gamma_{\ell+\ell'+1}q_{\ell j} \tilde{v}_{\ell'}
\bigr)
-\frac{b}{2}\mu_1(q_j w-p_j v)
\nonumber\\
&-\sum_{k} (-1)^k m(a_k)R_{jk}p_k.
\label{rs2}
\end{align}
By equating the right-hand sides of (\ref{ls1}) and (\ref{ls2}) 
with (\ref{rs1}) and (\ref{rs2}), respectively, 
the ``additional'' terms $D(\mathbb{I}-\mathbb{K}_I)^{-1}\varphi(a_j)$ and 
$D(\mathbb{I}-\mathbb{K}_I)^{-1}\psi(a_j)$
in (\ref{vars_del_j1}) and (\ref{vars_del_j2}) are expressed in terms of the quantities
$R_{jk}$, $q_{\ell j}$, $p_{\ell j}$, $u_\ell$, $w_\ell$, $v_\ell$, $\tilde{v}_\ell$,
leading to the nonuniversal equations [$(j,k)$ assumes (1,2) or (2,1)]:
\begin{align}
m(a_j) \frac{\d q_j}{\d a_j}&=
\sum_{\ell\geq 0}
\Bigl(
\alpha_\ell+\frac{b}{2}\mu_\ell +
\sum_{\ell'\geq 0}(\alpha_{\ell+\ell'+1}v_{\ell'}+\gamma_{\ell+\ell'+1}u_{\ell'})
-\frac{b}{2}\mu_1v\delta_{\ell,0}
\Bigr)q_{\ell j}
\nonumber\\
&+\sum_{\ell\geq 0}
\Bigl(
\beta_\ell +\sum_{\ell'\geq 0}(\alpha_{\ell+\ell'+1}u_{\ell'}+\beta_{\ell+\ell'+1}v_{\ell'})
+\frac{b}{2}\mu_1u\delta_{\ell,0}
\Bigr)p_{\ell j}
\nonumber\\
&-(-1)^k m(a_k)R_{jk}q_k,
\label{nu1a}\\
m(a_j) \frac{\d p_j}{\d a_j}
&=\sum_{\ell\geq 0}
\Bigl(
-\gamma_\ell +\sum_{\ell'\geq 0}(\alpha_{\ell+\ell'+1}w_{\ell'}+\gamma_{\ell+\ell'+1}\tilde{v}_{\ell'})
-\frac{b}{2}\mu_1w\delta_{\ell,0}
\Bigr)q_{\ell j}
\nonumber\\
&+\sum_{\ell\geq 0}
\Bigl(-\alpha_\ell+\frac{b}{2}\mu_\ell
+\sum_{\ell'\geq 0}(\alpha_{\ell+\ell'+1}\tilde{v}_{\ell'}+\beta_{\ell+\ell'+1}w_{\ell'})
+\frac{b}{2}\mu_1v\delta_{\ell,0}
\Bigr)p_{\ell j}
\nonumber\\
&-(-1)^k m(a_k)R_{jk}p_k.
\label{nu2a}
\end{align}
Note that the terms with $k=j$ in the sums $\sum_{k=1}^2$ of (\ref{rs1}) and (\ref{rs2})
are canceled by the first terms on the right-hand sides of (\ref{vars_del_j1}) and (\ref{vars_del_j2}).
Since the sums over $\ell, \ell'$ are finite,  
Eqs.\,(\ref{logdet}), (\ref{vars_der_k1}), (\ref{vars_der_k2}), (\ref{Rjk}), (\ref{Rkk}), (\ref{qlj}), (\ref{plj}), (\ref{nu1a}), and (\ref{nu2a})
form a closed system of coupled PDEs that determines $\det(\mathbb{I}-\mathbb{K}_I)$
as a function of $a_1$ and $a_2$.\footnote{
The last term within the curly braces in (\ref{MDKI4})
appeared in a more general form as Eq.\,(2.40) in \cite{Tracy:1994c}, 
though its implications (terms containing $(b/2)\mu_1\delta_{\ell, 0}$ in our linear case)
for the nonuniversal equations (\ref{nu1a}), (\ref{nu2a}) were not explicitly presented.
Moreover, an erroneous instruction was given on p.\,47 
to add $(b/2)\mu_\ell$ to the second parenthesis
in (\ref{nu1a}) and to the first parenthesis in (\ref{nu2a}).}

\section{Tracy-Widom system and probability distributions} 
\subsection{TW system for J\'{a}nossy density}
By substituting the coefficients $\alpha_\ell$, etc.~in (\ref{coeffs}) and $b=i$
into the TW system  listed just above,
and eliminating $q_{1j}$ and $p_{1j}$ in favor of  $q_{j}$ and $p_{j}$,
the set of PDEs that determines the J\'{a}nossy density $\tilde{J}_1(0;[a_1,a_2])$
of CUE$_N$ reads [$(j,k)$ assumes (1,2) or (2,1)]:
\begin{align}
i(1-\e^{i a_j})
\frac{\d q_j}{\d a_j}
&=-\left\{
\frac{N+1}{2} (1-\e^{i a_j})+v-\frac{1}{N}
\right\}q_j+
(N+1) \left(u-\frac1N\right)p_j
\nonumber\\
&~~~-(-1)^k i(1-\e^{i a_k}) R_{jk}q_k,
\nonumber\\
i(1-\e^{i a_j})\frac{\d p_j}{\d a_j}
&=\left\{
\frac{N-1}{2} (1-\e^{i a_j})+v-\frac{1}{N}
\right\}p_j
+(N-1)\left(w-\frac1N\right)q_j
\nonumber\\
&~~~-(-1)^k i(1-\e^{i a_k})R_{jk}p_k,
\nonumber\\
\frac{\d q_j}{\d a_k}&=(-1)^k R_{jk} q_k,~~
\frac{\d p_j}{\d a_k}=(-1)^k R_{jk} p_k,
\nonumber\\
\frac{\d u}{\d a_j}&=(-1)^j q_j^2,~~
\frac{\d w}{\d a_j} =(-1)^j p_j^2,~~
\frac{\d v}{\d a_j} =(-1)^j q_j p_j,
\nonumber\\
\frac{\d}{\d a_j}&\log\tilde{J}_1(0;[a_1,a_2])=-(-1)^{j} R_{jj},
\label{TWsystem}
\end{align}
where ${\displaystyle R_{jk}=\frac{q_j p_k-p_j q_k}{\e^{i a_j}-\e^{i a_k}}}$ and 
${\displaystyle R_{jj}=
-i \,\e^{-i a_j}\Bigl(p_j\frac{\d q_j}{\d a_j}-q_j\frac{\d p_j}{\d a_j}\Bigr)}$.
The boundary conditions at $a_1, a_2 \to 0$ follow from 
the Neumann expansion of the definitions (\ref{FredholmDet}), (\ref{vars}):
\begin{align}
&q_j=i\frac{N+1}{2\sqrt{2\pi}}a_j-\frac{(N+1)(N+2)}{12\sqrt{2\pi}}a_j^2
-i\frac{(N+1)^3}{48\sqrt{2\pi}}a_j^3+\cdots,
\nonumber\\
&p_j=-i\frac{N-1}{2\sqrt{2\pi}}a_j-\frac{(N-1)(N-2)}{12\sqrt{2\pi}}a_j^2
+i\frac{(N-1)^3}{48\sqrt{2\pi}}a_j^3+\cdots,
\nonumber\\
&u=\frac{(N+1)^2}{24\pi}(a_1^3-a_2^3)+\cdots,~
w=\frac{(N-1)^2}{24\pi}(a_1^3-a_2^3)+\cdots,~
v=-\frac{N^2-1}{24\pi}(a_1^3-a_2^3)+\cdots,
\nonumber\\
&\log\tilde{J}_1(0;[a_1,a_2])=\frac{N(N^2-1)}{72\pi}(a_1^3-a_2^3)+\cdots,
\label{bc}
\end{align}
up to terms of $\order{a_1, a_2}^4$.
Note that, after unfolding of variables $a_j\mapsto (2/N)a_j$, the TW system
(\ref{TWsystem}), (\ref{bc}) is invariant under the sign flip $N\mapsto -N$,
accompanied by the notation changes
$q_j \leftrightarrow \overline{p_j},  
u \leftrightarrow -\overline{w}, v \mapsto -\overline{v}.$
Accordingly, the J\'{a}nossy density in the unfolded variables,
along with other probability distributions subsequently derived from it,
are all functions of $1/N^2$ that are analytic at $1/N^2=0$, as required by (\ref{Ksin}).

For the numerical evaluation of $\tilde{J}_1(0;[a,b])$,
the simplest approach is to start from the above boundary conditions at
$(\epsilon a, \epsilon b)$ with a sufficiently small $\epsilon>0$ and integrate
the TW system (\ref{TWsystem}) along $\left(a_1(s), a_2(s))=(sa, sb\right)$
with respect to $s$, $\epsilon \leq s \leq 1$, using 
${\frac{d}{ds}=a \frac{\d}{\d a_1}+b\frac{\d}{\d a_2}}$.
As a cross-check, we have verified that the values of $J_1(0;[a_1,a_2])$ 
obtained via this prescription agree precisely with those from the Nystr\"{o}m-type
quadrature approximation of the Fredholm determinant \cite{Bornemann:2010a}
\begin{align}
\det\bigl(\mathbb{I}-\tilde{\mathbb{K}}_I\bigr)
\approx \det\left[ \delta_{ij} -\tilde{K}(x_i ,x_j)\sqrt{w_i w_j} \right]_{i,j=1}^m~.
\label{Nystrom}
\end{align}
Here $\{x_i, w_i\}_{i=1}^m$ is the $m$-th order quadrature of the interval $I$ such that 
$\sum_{i=1}^m w_i f(x_i)\stackrel{m\to\infty}{\longrightarrow}\int_I dx\,f(x)$,
with a sufficiently large $m$.
Specifically, for $N=10$, the relative deviations of $\tilde{J}_1(0;[a_1,a_2])$
between values computed from the TW system starting from the initial value $\epsilon=10^{-15}$
using Mathematica's {\tt NDSolve}
with {\tt WorkingPrecision}\,$\to$\,5\,{\tt MachinePrecision}
and those from the Nystr\"{o}m-type approximation (\ref{Nystrom}) with
Gauss-Legendre quadrature of order $m=256$ are on the orders of 
${10^{-36}}$ for $|a_1|, a_2\leq \Delta=2\pi/N$ (the mean level spacing),
of ${10^{-31}}$ for $|a_j|\leq 2\Delta$, and
of ${10^{-11}}$ for $|a_j|\leq 3\Delta$.
A Mathematica notebook {\tt J1CUE.nb} taylored for these enumerations
is included in Supplementary materials.

In the large-$N$ limit after unfolding $x\mapsto (2/N)x$,
$\tilde{\varphi}(x)$ and $\tilde{\psi}(x)$ in (\ref{Ktil})
form a complex conjugate pair, implying
$\overline{q_j}=p_j$ and $\overline{u}=w$ according to the definitions (\ref{vars}).
Accordingly, the second, non-universal equation in (\ref{TWsystem})
reduces to the complex conjugate of the first, which now takes the form
(after redefining $a_j\mapsto (2/N)a_j$ and $u\mapsto u/N$):
\begin{align}
a_j \frac{\d q_j}{\d a_j}
= i a_j q_j+(u-1) \overline{q_j}-(-1)^k a_k R_{jk}q_k,~~
R_{jk}=\frac{q_j \overline{q_k}-\overline{q_j} q_k}{a_j-a_k}.
\label{TWsin}
\end{align}
This is equivalent to the second and the third of Eq.\,(13) in \cite{Nishigaki:2024}
for the sine kernel case, after  changes in notation
$\mathrm{Re}(q_j)\mapsto p_j$, $\mathrm{Im}(q_j)\mapsto q_j$,
$\mathrm{Re}(u)\mapsto w-u$, $\mathrm{Im}(u)\mapsto 2v$.\footnote{
These notational conventions correspond to
different decompositions of the numerator of the sine kernel,
$\sin(x-y)=\sin x \cos y-\cos x \sin y=(-i/2) (\e^{ix} \e^{-iy}-\e^{-ix}\e^{iy})$,
related by a global $\mathrm{Sp}(2)$ transformation
acting on $\Phi(x)$ and $\tilde{\Phi}(x)$.}

\subsection{Symmetric interval case}
The number of independent and dependent variables reduces
when the interval is symmetric about the origin, $I=[-t,t]$;
then the properties
$\overline{\varphi(x)}=\varphi(-x)$, $\overline{\psi(x)}=\psi(-x)$ imply that
\begin{align}
q(t):=q_2=\overline{q_1},~~
p(t):=p_2=\overline{p_1},~~
u(t), w(t), v(t) \in \mathbb{R}.
\nonumber
\end{align}
By using 
${':=\frac{d}{dt}=\frac{\d}{\d a_2}-\frac{\d}{\d a_1}}$,
the set of PDEs (\ref{TWsystem})
combines to ordinary differential equations (ODEs) of the form:
\begin{align}
&i(1-\e^{it})q'=-
\left\{
\frac{N+1}{2} (1-\e^{i t})+v-\frac{1}{N}
\right\}q+
(N+1) \left(u-\frac1N\right)p
+\tan \frac{t}{2}\,(q \bar{p}-p \bar{q})\,\bar{q},
\nonumber\\
&i(1-\e^{it})p'=
\left\{
\frac{N-1}{2} (1-\e^{i t})+v-\frac{1}{N}
\right\}p
+(N-1)\left(w-\frac1N\right)q
+\tan \frac{t}{2}\,(q \bar{p}-p \bar{q})\,\bar{p},
\nonumber\\
&u'=q^2+\bar{q}^2,~~
w'=p^2+\bar{p}^2,~~
v'=qp+\bar{q}\bar{p},~~
\nonumber\\
&\Bigl(\log\tilde{J}_1(0;[-t,t])\Bigr)'
=i\,\e^{-it}(pq'-qp')-i\,\e^{it}(\bar{p}\bar{q}'-\bar{q}\bar{p}')
+\cot t \,(q\bar{p}-p\bar{q})^2,
\label{TWsystem_t}
\end{align}
with boundary conditions (\ref{bc}) at $a_2=-a_1=t$.

As mentioned in the previous subsection, $\overline{q}=p$ and $u=w\in \mathbb{R}$ hold
in the large-$N$ limit after unfolding $t\mapsto (2/N)t$.
Then the system of ODEs (\ref{TWsystem_t}) takes a simpler form
(after redefining $u\mapsto u/N$):
\begin{align}
q'=i q+\frac{u-1}{t}\bar{q},~~
u'=2 (q^2+\bar{q}^2),~~
\Bigl(\log\tilde{J}_1\Bigr)'=2i(\bar{q}q'-q \bar{q}')
+\frac{(q^2-\bar{q}^2)^2}{t}.
\label{FO}
\end{align}
This set of ODEs was first derived in \cite{Forrester:1996} as its Eqs.\,(13)--(17),
with changes in notation
$\mathrm{Re}(q)\mapsto q$,  $\mathrm{Im}(q)\mapsto -p$, $u\mapsto u-w.{}^\S$
Ref.\,\cite{Forrester:1996} also observed that $u$ can be eliminated from (\ref{FO})
and its conjugate, resulting in $q(t)$ satisfying an autonomous nonlinear DE:
\begin{align}
q'=iq+\frac{2\mathrm{Im}(q^2)-1}{t}\bar{q},~~
q(t)=\frac{i t}{\sqrt{2\pi}} - \frac{t^2}{3\sqrt{2\pi}}-\frac{i t^3}{6\sqrt{2\pi}}+\cdots.
\nonumber
\end{align}

\subsection{Nearest-neighbor spacing distribution}
A variety of spectral-statistical distributions of interest 
can be derived from the J\'{a}nossy density in (\ref{TWsystem}).
The difference 
$\tilde{J}_1(0;[-t,t])-\tilde{J}_1(0;[-t-dt,t+dt])$ represents the probability that
the spacing between an eigenvalue (assumed to be at the origin) and its
{\em nearest neighbor}, either to the left or right,
lies within the interval $[t, t+dt]$.
Equivalently, the distribution $P_\mathrm{nn}(t)$ of
the nearest neighbor spacing $t=\min(|a_1|, a_2)$ is given by \cite{Forrester:1996}
\begin{align}
P_\mathrm{nn}(t)=-\frac{d\tilde{J}_1(0;[-t,t])}{dt}.
\label{Pnn}
\end{align}
The plots of $P_\mathrm{nn}(t)$
for various $N$
and their deviations from the large-$N$ (sine-kernel) limit 
$P_\mathrm{nn}^{(0)}(t)$ derived from (\ref{FO}), scaled by $N^2$,
are exhibited in Fig.\,1.
The convergence of all curves in the right panel indicates that the leading $\order{N^{-2}}$ 
correction admits a well-defined limit,
\[
P_\mathrm{nn}(t)=
P_\mathrm{nn}^{(0)}(t)
+\frac{1}{N^2}P^{(2)}_\mathrm{nn}(t)+\mathcal{O}\Bigl(\frac{1}{N^{4}}\Bigr).
\]
These plots of $P_\mathrm{nn}(t)$ and its deviation from the sine-kernel limit
was first presented in [5, Fig.\,7], which instead employed enumeration via
the Nystr\"{o}m-type approximation (\ref{Nystrom}).
\begin{figure}[b] 
\begin{center}
\includegraphics[bb=0 0 360 235,width=75mm]{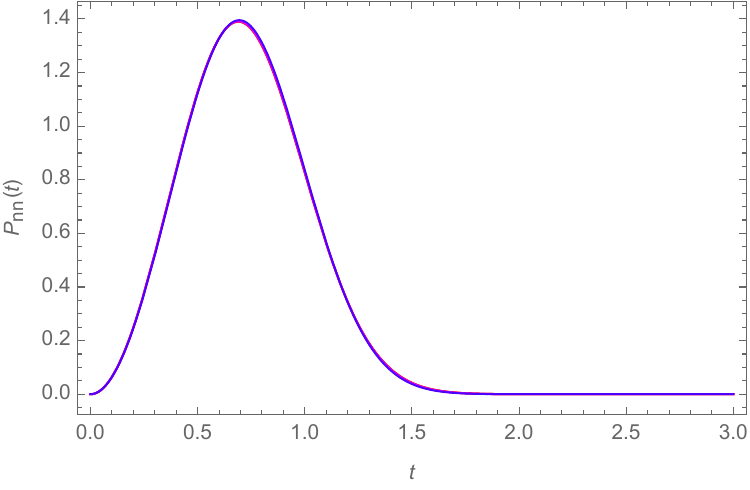}~
\includegraphics[bb=0 0 360 235,width=75mm]{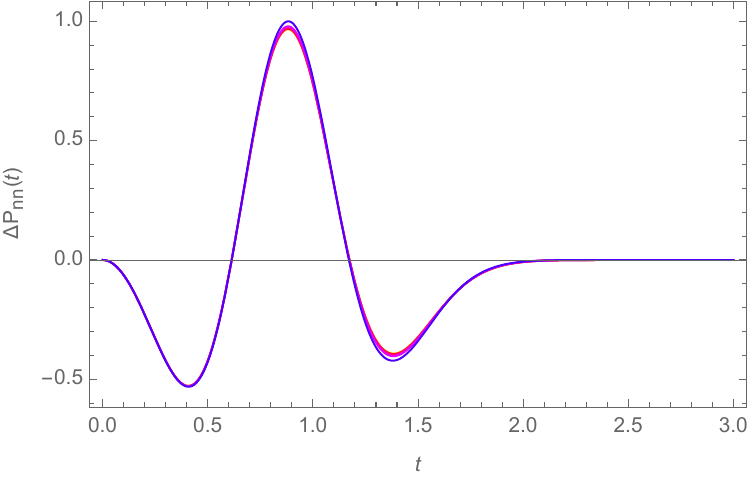}
\caption{
Nearest neighbor eigenvalue spacing distributions $P_{\mathrm{nn}}(t)$ of CUE$_N$
for $N=8, 12, 16, 20, 24$ (blue to red) [left]
and their deviations from the sine-kernel limit scaled by $N^2$,
$\Delta P_\mathrm{nn}(t)= 
N^2(P_{\mathrm{nn}}(t)-P^{(0)}_{\mathrm{nn}}(t))$ [right].
For visual clarity, eigenvalue variables for CUE$_N$ in Figs.\,1, 2, and 4
are rescaled (unfolded) by a factor $N/2\pi$ relative to the main text,
so that the mean eigenvalue spacing is normalized to unity.
}
\end{center}
\end{figure}

\subsection{Gap-ratio distribution}
For the same reason that Eq.\,(\ref{Pnn}) holds,
the joint distribution $P_{\mathrm{c}}(a_1, a_2)$
of two consecutive eigenvalue spacings $-a_1$ and $a_2$ is given by
\begin{align}
P_{\mathrm{c}}(a_1,a_2)=
-\frac{\d^2 \tilde{J}_1(0;[a_1,a_2]) }{\d a_1 \d a_2},
\label{Pc}
\end{align}
from which the distribution $P_{\mathrm{r}}(r)$ of the gap ratio $r=|a_1|/a_2$
follows:
\begin{align}
P_{\mathrm{r}}(r)&=
\int\!\!\!\!\int_{a_1<0<a_2<2\pi+a_1}
\!\!\!\!\!\!\!\!\!\!\!\!\!\!\!\!\!\!\!\!\!\!\!\!\!\!\!\!\!\!
da_1 da_2\,
P_{\mathrm{c}}(a_1,a_2)\,
\delta(r-|a_1|/a_2)
=\int_0^{2\pi/(1+r)}
\!\!\!\!\!\!\!\!\!\!\!\!\!\!\!
db\, b\, P_{\mathrm{c}}(-r b, b).
\label{Prr}
\end{align}
Upon changing the variable from $r$ to 
$\tilde{r}:=\min(|a_1|, a_2)/\max(|a_1|, a_2)=\min(r, r^{-1})\in [0,1]$,
the corresponding distribution is given by $2P_{\mathrm{r}}(\tilde{r})$.
As $N$ increases, both $P_{\mathrm{c}}(a,b)$ and $P_{\mathrm{r}}(r)$
converge to their respective sine-kernel limits,
$P^{(0)}_{\mathrm{c}}(a,b)$ and $P^{(0)}_{\mathrm{r}}(r)$,
derived from (\ref{TWsin}) (Figs.\,2 and 3, left panels).
\begin{figure}[b] 
\begin{center}
\includegraphics[bb=0 0 360 224,width=76mm]{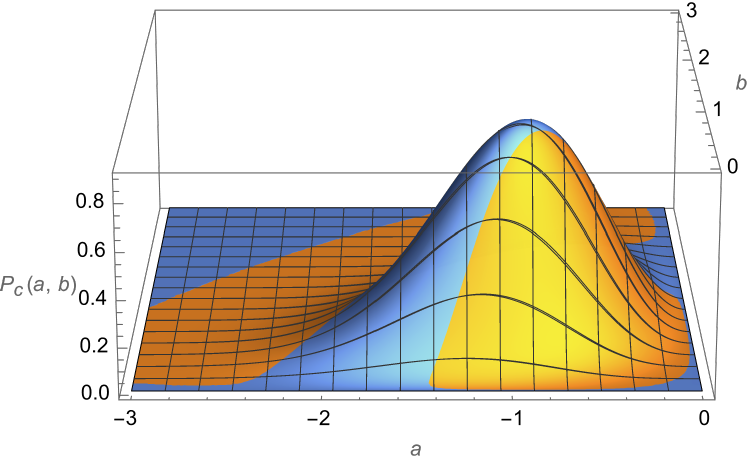}~
\includegraphics[bb=0 0 360 244,width=70mm]{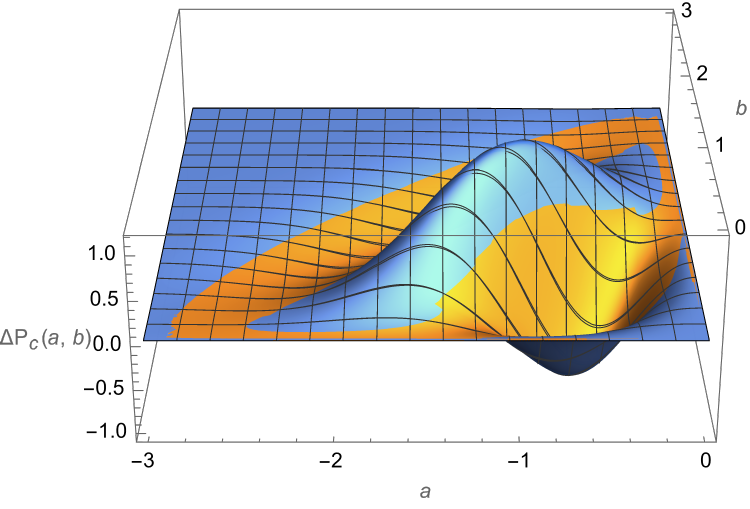}
\caption{
Joint distributions of two consecutive eigenvalue spacings
$P_{\mathrm{c}}(a,b)$ of CUE$_N$ for $N=8, 16$ (blue, orange) [left]
and their deviations from the sine-kernel limit scaled by $N^2$,
$\Delta P_{\mathrm{c}}(a,b)=N^2(P_{\mathrm{c}}(a,b)-
P^{(0)}_{\mathrm{c}}(a,b))$ [right].
}
\end{center}
\end{figure}
\begin{figure}[b] 
\begin{center}
\includegraphics[bb=0 0 360 235,width=75mm]{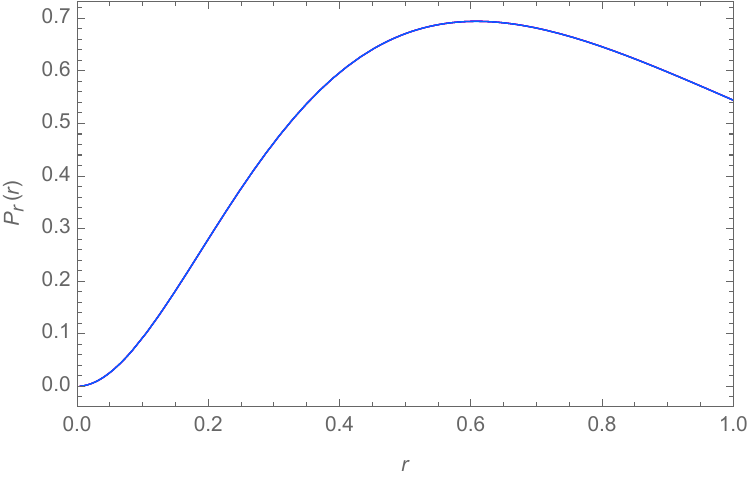}~
\includegraphics[bb=0 0 360 235,width=75mm]{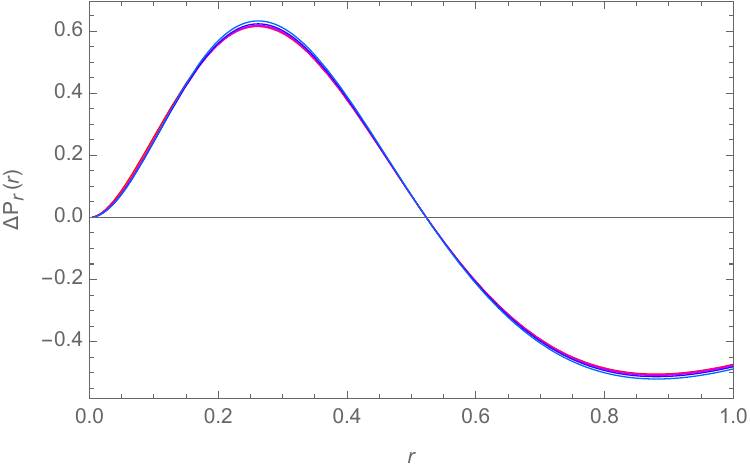}
\caption{
Gap-ratio distributions $P_{\mathrm{r}}(r)$ of CUE$_N$
for $N=8, 10, 12, 14, 16$ (blue to red) [left]
and their deviations from the sine-kernel limit scaled by $N^4$,
$\Delta P_{\mathrm{r}}(r)=N^4(P_{\mathrm{r}}(r)-P^{(0)}_{\mathrm{r}}(r))$
[right].
}
\end{center}
\end{figure}
Remarkably, however, the orders of their deviations differ:
while the leading correction to $P^{\mathrm{(0)}}_{\mathrm{c}}(a,b)$ 
exhibits a scaling limit when multiplied by
$N^2$ as expected from (\ref{Ksin}) and evident in Fig.\,2 (right panel),
the leading correction to $P^{\mathrm{(0)}}_{\mathrm{r}}(r)$,
contrary to na\"{\i}ve expectations,
displays a scaling limit only when multiplied by $N^4$,
as clearly shown in Fig.\,3 (right panel).
In other words, the vanishing of the would-be leading correction $P^{(2)}_\mathrm{r}(t)=0$
in the $1/N^2$ expansion of the gap-ratio distribution of CUE$_N$ for large $N$,
\[
P_\mathrm{r}(r)=
P_\mathrm{r}^{(0)}(r)
+\frac{1}{N^2}P^{(2)}_\mathrm{r}(r)
+\frac{1}{N^4}P^{(4)}_\mathrm{r}(r)
+\mathcal{O}\Bigl(\frac{1}{N^{6}}\Bigr),
\]
enables direct access to the next-to-leading correction $P^{(4)}_\mathrm{r}(t)$,
which captures the $\order{N^{-4}}$ contribution of the kernel (\ref{Ksin}).
Nevertheless, the mechanism underlying the intricate cancellation of
$\order{N^{-2}}$ contributions in the correlated fluctuations of consecutive spacings
remains obscure from the structure of 
the PDEs (\ref{TWsystem}) and the boundary conditions (\ref{bc}).

Our findings suggest the potential to extract subtle spectral
information from quantum-chaotic systems that would otherwise be inaccessible:
Suppose one is given a spectrum of a DPP of $N$ energy levels,
with an unknown underlying kernel, but whose correlations converge
asymptotically to those of the sine kernel as $N \to \infty$.
Agreement in the deviation $P_{\mathrm{DPP,c}}(a,b)-P^{(0)}_{\mathrm{c}}(a,b)$
(or in the deviations of other observables, 
such as the pair correlation function or spacing distribution)
with those of CUE$_N$ indicates that $K_{\mathrm{DPP}}(x,y)$ coincides with the CUE$_N$
kernel at $\order{N^{-2}}$.
In contrast, agreement in the deviation $P_{\mathrm{DPP,c}}(r)-P^{(0)}_{\mathrm{c}}(r)$
suggests an even closer match, extending to $\order{N^{-4}}$.
In the next section,
we leverage this advantage of the probability distribution of gap ratios $r=a/b$
to conduct a statistical analysis of the nontrivial zeros of the Riemann zeta function.

\section{Riemann zeta zeros}
\subsection{Joint distribution of two consecutive spacings}
Following the conditional proof \cite{Rudnick:1996} of the Montgomery-Odlyzko law
\cite{Montgomery:1972,Odlyzko:1987}
which asserts that the statistical correlations of the zeros of principal 
$L$-functions $\{1/2\pm i \gamma_n\}$ asymptotically form a DPP governed by the sine kernel,
and Odlyzko's discovery \cite{Odlyzko:2001} of a systematic structure
in the deviation of the spacing distribution from the sine-kernel limit,
it is natural to investigate how the finite-size corrections on the Riemann side
compare with those of CUE$_N$ \cite{Bogomolny:2006,Forrester:2015,Bornemann:2017}.
There, both theoretical and numerical evidence have been presented indicating
that the unfolded Riemann zeta zeros around $\gamma_n\approx T\gg 1$
exhibit the behavior of a DPP governed by the kernel
\begin{align}
K_{\mathrm{RZ}}(x,y)=
\frac{\sin(x-y)}{\pi(x-y)}+\frac{(x-y) \sin (x-y)}{6 \pi \,N_{\mathrm{e}}^2}
+\frac{Q}{\sqrt{3}\Lambda^{3/2}}
\frac{(x-y)^2 \cos (x-y)}{6 \pi \,N_{\mathrm{e}}^3}
+\mathcal{O}\Bigl(\frac{1}{N_{\mathrm{e}}^{4}}\Bigr)
\label{KRZ}
\end{align}
under an identification of ``$N$ effective",
\begin{align}
N_{\mathrm{e}}(T)=\frac{1}{\sqrt{12\Lambda}}\log\frac{T}{2\pi},
\label{Neff}
\end{align}
deduced from matching of the pair correlation function $R_2(x,y)$
\cite{Bogomolny:2006}.
Here, the arithmetic constants $\Lambda=1.573151071\ldots$ and $Q=2.315846384\ldots$
arise from prime sums.

We used Odlyzko's high-precision dataset (announced in \cite{Odlyzko:2001})
of $1041208320$ Riemann zeta zeros starting
from $\gamma_n=13066434408793621120027.3961\ldots$,
corresponding to $n=1.00000000000000985531550 \times 10^{23}$.
Fig.\,4 displays the joint distribution $P_\mathrm{RZ, c}(a,b)$
of two unfolded consecutive spacings $a,b=\rho(\gamma_n)(\gamma_{n\mp 1}-\gamma_n)$, 
with $\rho(T)=\frac{1}{2\pi} \log\frac{T}{2\pi}$
representing the asymptotic density of zeros,
along with its deviation from the sine-kernel limit scaled by 
$N_{\mathrm{e}}^2=11.2975909009\ldots^2$.
The excellent agreement between the Riemann zero data and the analytic predictions of
CUE$_{N_{\mathrm{e}}}$ in both panels provides further confirmation of the empirical
formula (\ref{KRZ}), (\ref{Neff}) at $\order{N_{\mathrm{e}}^{-2}}$,
complementing previous observations based on 
the spacing distribution $p(s)$ \cite{Bogomolny:2006},
the pair correlation function $R_2(x,y)$ \cite{Forrester:2015},
and the nearest-neighbor spacing distribution $P_{\mathrm{nn}}(t)$ \cite{Bornemann:2017}.
\begin{figure}[t] 
\begin{center}
\includegraphics[bb=0 0 450 293,width=75mm]{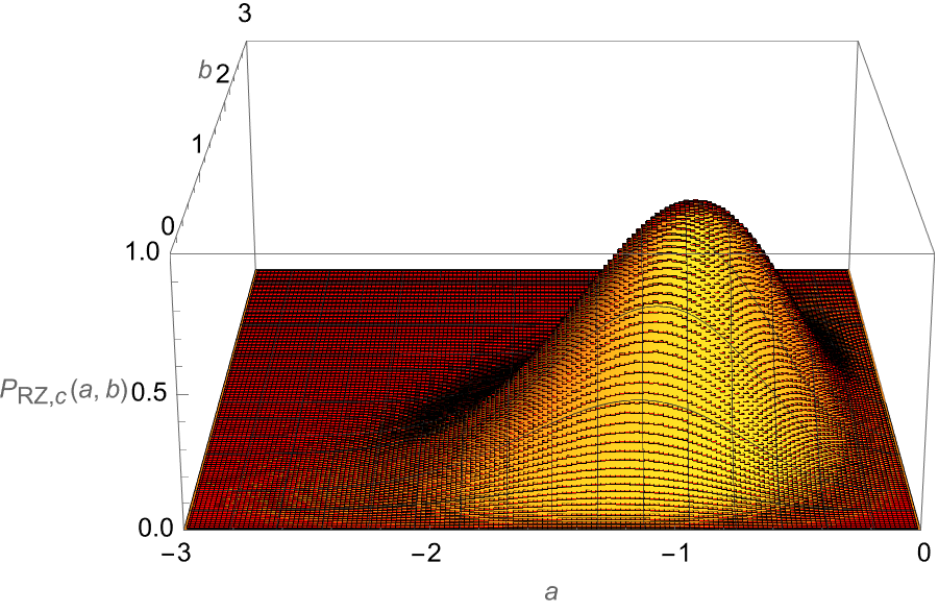}~
\includegraphics[bb=0 0 450 288,width=76mm]{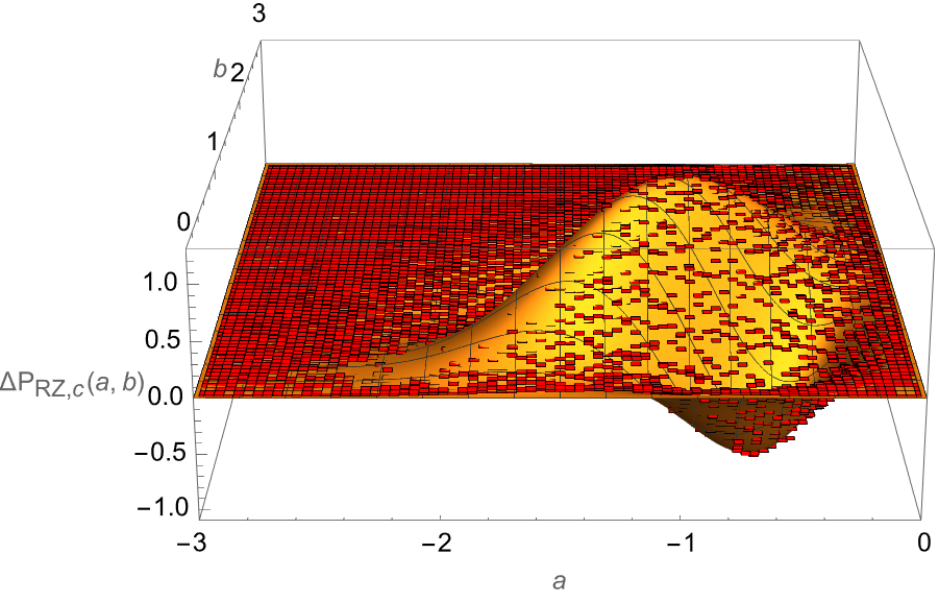}
\caption{
Joint distribution $P_\mathrm{RZ, c}(a,b)$ of two unfolded adjacent spacings
of $10^9$ Riemann zeta zeros $\{\gamma_n\}$ around $n= 1.000\times 10^{23}$ [left]
and its deviation from the sine-kernel limit scaled by $N_{\mathrm{e}}^2$,
$\Delta P_{\mathrm{RZ, c}}(a,b)=
N_{\mathrm{e}}^2(P_{\mathrm{RZ, c}}(a,b)-P^{(0)}_{\mathrm{c}}(a,b))$ [right].
Each histogram (red) is overlayed with its counterpart for CUE$_{N_{\mathrm{e}}}$ (orange).}
\end{center}
\end{figure}
\subsection{Gap-ratio distribution}
Given the observation in the previous section that the would-be leading $\order{N^{-2}}$
correction in the gap-ratio distribution of CUE$_N$ is absent,
and noting that Eq.\,(\ref{KRZ}) agrees with CUE$_{N_{\mathrm{e}}}$ (\ref{Ksin})
at $\order{N_{\mathrm{e}}^{-2}}$ but differs at $\order{N_{\mathrm{e}}^{-3}}$,
it is natural to anticipate that the gap-ratio distribution of the Riemann zeros admits
a large-$N_{\mathrm{e}}$ expansion of the form
\[
P_\mathrm{RZ, r}(r)=
P_\mathrm{r}^{(0)}(r)
+\frac{1}{N_{\mathrm{e}}^3}P^{(3)}_\mathrm{RZ, r}(r)
+\mathcal{O}\Bigl(\frac{1}{N_{\mathrm{e}}^{4}}\Bigr) .
\]
In Fig.\,5 we plot histograms $P_\mathrm{RZ, r}(r)$ of gap ratios 
$r=(\gamma_{n+1}-\gamma_n)/(\gamma_{n}-\gamma_{n-1})$
of $10^8$ Riemann zeros starting from 
$\gamma_n=30581878184.0869433888221\ldots$
\cite{LMFDB} and of $\approx 10^9$ Riemann zeros starting from 
$2513274122880031.43550662\ldots$ and 
$13066434408793621120027.3\ldots$, 
corresponding to 
$n=1.03700788360\times 10^{11}, 1.30489942652\ldots \times 10^{16},  
1.00000000000\ldots \times 10^{23}$,
and their deviations of from the sine-kernel limit $P^{(0)}_{\mathrm{r}}(r)$
\cite{Nishigaki:2024} scaled by 
$N_{\mathrm{e}}^3=5.13383486853\ldots^3, 7.73844996441\ldots^3, 11.2975909009\ldots^3$,
respectively.
\begin{figure}[t] 
\begin{center}
\includegraphics[bb=0 0 360 231,width=75mm]{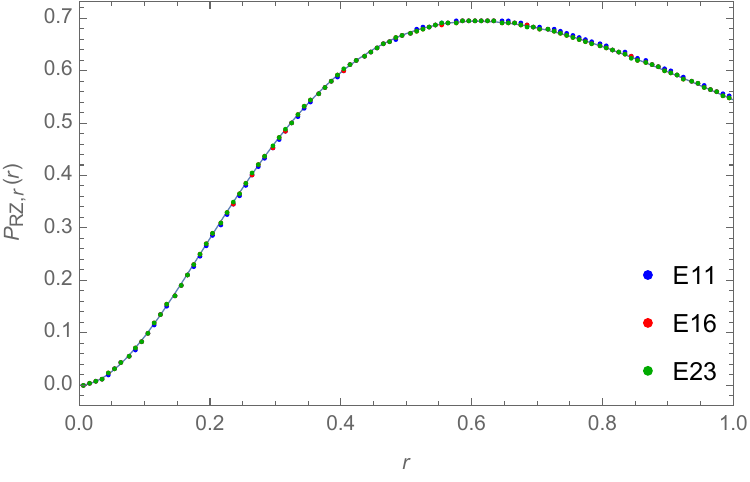}~
\includegraphics[bb=0 0 360 227,width=75.5mm]{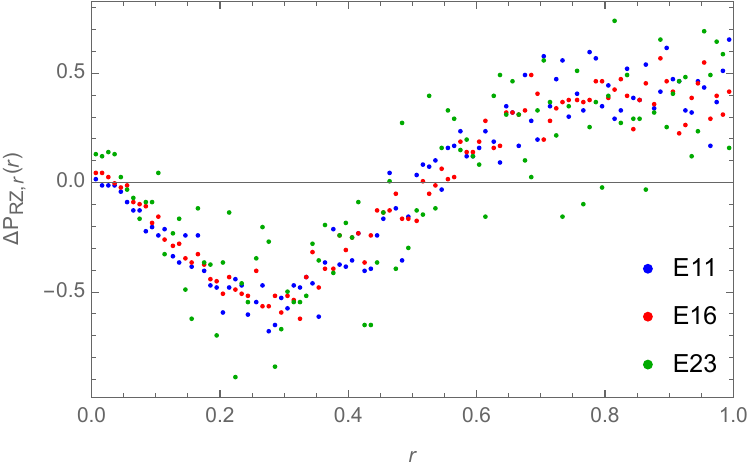}
\caption{
Gap-ratio distributions $P_{\mathrm{RZ, r}}(r)$ of the Riemann zeta zeros $\{\gamma_n\}$
around $n= 1.037\times 10^{11}, 1.304\times 10^{16}, 1.000\times 10^{23}$
(blue, red, green dots) and of the sine kernel $P^{(0)}_{\mathrm{r}}(r)$
(curve) [left], and their deviations scaled by $N_{\mathrm{e}}^3$,
$\Delta P_{\mathrm{RZ, r}}(r)=
N_{\mathrm{e}}^3(P_{\mathrm{RZ, r}}(r)-P^{(0)}_{\mathrm{r}}(r))$ [right].}
\vspace{2mm}
\includegraphics[bb=0 0 450 293,width=75mm]{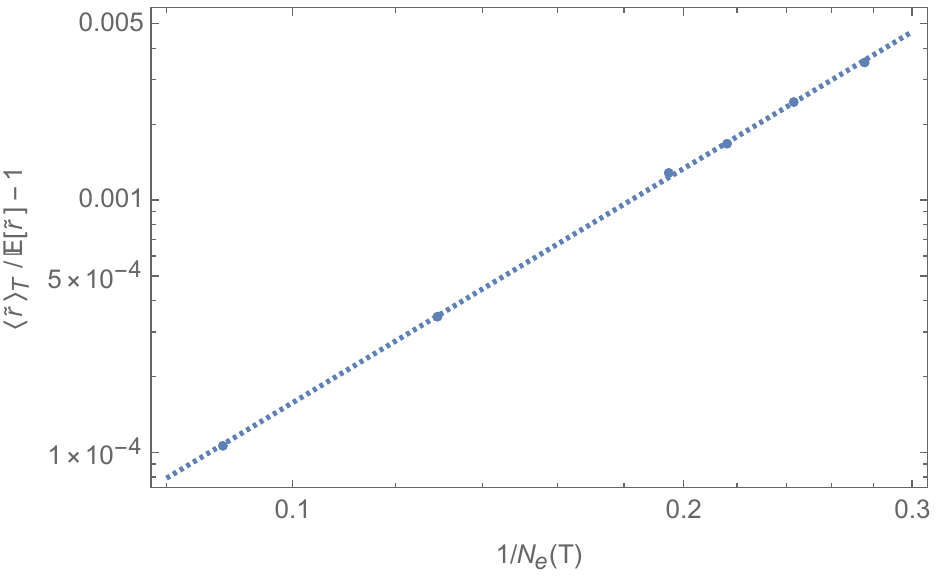}
\caption{
Relative deviations of the mean gap ratios $\langle\tilde{r}\rangle_T$
of the Riemann zeta zeros $\{\gamma_n\}\approx T$, computed in windows around
$n= 10^8, 10^9, 10^{10}, 1.037\times 10^{11}, 1.304\times 10^{16}, 1.000\times 10^{23}$,
from the sine-kernel limit.
The data are shown on a log-log plot against $N_{\mathrm{e}}(T)^{-1}$.
The dotted line represents the optimal linear fit to the six data points,
$0.1896 \,N_{\mathrm{e}}^{-3.081}$.
}
\end{center}
\end{figure}
Since our focus is on the higher-order $\order{N_{\mathrm{e}}^{-3}}$
correction to the sine-kernel limit rather than the leading $\order{N_{\mathrm{e}}^{-2}}$
correction analyzed in \cite{Bogomolny:2006,Forrester:2015,Bornemann:2017},
significant fluctuations in these histograms are inevitable.
Nevertheless, despite these fluctuations, 
Fig.\,5  (right panel) reveals that the histograms exhibit a limiting structure\footnote{
Because the conjectured kernel (\ref{KRZ}) can be adjusted to match the
$1/N^2$-expansion of the CUE$_N$ kernel (\ref{Ksin}) up to $\order{N^{-4}_{\mathrm{e}}}$,
\[
K_{\mathrm{RZ}}(x,y)=
\frac{\sin(x-y)}{\pi(x-y)}
+\frac{(x-y) \sin \left(\bar{\alpha}(x-y)\right)}{6 \pi \,N_{\mathrm{e}}^2}
+\mathcal{O}\Bigl(\frac{1}{N_{\mathrm{e}}^{4}}\Bigr),
\]
by incorporating the $\order{N_{\mathrm{e}}^{-3}}$ term with
$\bar{\alpha}=1+Q/(\sqrt{3}\Lambda^{3/2}N_{\mathrm{e}})$ \cite{Bogomolny:2006,Bornemann:2017},
the limiting function that emerges in Fig.\,5 (right) as $N_{\mathrm{e}}\to\infty$ 
can still be plausibly attributed to a ``random matrix origin".}
consistent with $P^{(2)}_\mathrm{RZ, r}(r)=0$,
in agreement with CUE$_N$.
The $N_{\mathrm{e}}^{-3}$ scaling of the finite-size correction
in the mean value of gap ratios $\tilde{r}=\min(r, r^{-1})$
for the Riemann zeros, relative to the sine-kernel prediction
$\mathbb{E}[\tilde{r}]=2\int_0^1 dr\,r\,P^{(0)}_{\mathrm{r}}(r)=0.5997504209\ldots$
was reported in our previous work [13, Table 1 and Fig.\,4] and is updated here in Fig.\,6.
The origin of the exponent $-3$, which was unclear at an earlier stage of
our investigation, is now elucidated.

\section*{Supplementary materials}
{\tt J1CUE.nb}:
Mathematica notebook for generating $\tilde{J}_{1}(0;[a, b])$ and $P_{\mathrm{c}}(a, b)$
analytically by the Tracy-Widom system (\ref{TWsystem})--(\ref{bc})
or numerically by the Nystr\"{o}m-type approximation (\ref{Nystrom}).

\section*{Funding}
The research of SMN is supported by Japan Society for the Promotion of Sciences
(JSPS) Grants-in-Aids for Scientific Research (C) No.\,7K05416.

\section*{Acknowledgments}
I am grateful to Peter J. Forrester for his valuable comments on the manuscript,
for insightful discussions during my visit to the University of Melbourne
(Feb.--Mar. 2025), and for the financial support that made the visit possible.
My thanks extend as well to all members of his group for their warm hospitality.

\end{document}